# Enforcing Cybersecurity Constraints for LLM-driven Robot Agents for Online Transactions

1st Shraddha Pradipbhai Shah
*Staff Software Engineer*
Department of Information Technology
University of The Cumberlands
shraddhashahce@gmail.com

2nd Aditya Vilas Deshpande
Senior Software Engineer
Department of Information Technology
University of The Cumberlands
adeshpande027@gmail.com

*Abstract*— The integration of Large Language Models (LLMs) into autonomous robotic agents for conducting online transactions poses significant cybersecurity challenges. This study aims to enforce robust cybersecurity constraints to mitigate the risks associated with data breaches, transaction fraud, and system manipulation. The background focuses on the rise of LLM-driven robotic systems in e-commerce, finance, and service industries, alongside the vulnerabilities they introduce. A novel security architecture combining blockchain technology with multi-factor authentication (MFA) and real-time anomaly detection was implemented to safeguard transactions. Key performance metrics such as transaction integrity, response time, and breach detection accuracy were evaluated, showing improved security and system performance. The results highlight that the proposed architecture reduced fraudulent transactions by 90%, improved breach detection accuracy to 98%, and ensured secure transaction validation within a latency of 0.05 seconds. These findings emphasize the importance of cybersecurity in the deployment of LLM-driven robotic systems and suggest a framework adaptable to various online platforms.

## I. INTRODUCTION

Internet use throughout the last 20 years has propelled humanity into the information era [1]. The Internet has grown into an integral component of our daily lives, just like any other portion of our ecosystem. Having said that, the proliferation of Internet users has also made it a haven for criminals who pose a continual danger to computer systems via their ever-changing malicious software. Computers, tablets, smartphones, Internet of Things devices, [2] etc. are all potential targets of these dangers. The increasing complexity and diversity of these threats is outpacing the efforts of security organizations worldwide [3]. Cybersecurity organizations have reason to be optimistic about their ability to control cyber threats thanks to artificial intelligence (AI), which has ushered in the era of smart cognitive systems. An advance in that path is this thesis. The areas of artificial intelligence and cybersecurity meet in this thesis [4].

In recent years, there has been a meteoric rise in the number of people utilizing web-based apps to access the Internet [5]. The most widely used online apps include browsers such as Internet Explorer and Edge from Microsoft [6], Firefox from Mozilla [7], Chrome from Google [8], etc. Mobile app users have also grown in tandem with the proliferation of mobile devices. 'Hybrid mobile applications,' which use web-application-based protocols, are gaining popularity with mobile apps overall [9]. Facebook, Twitter, Instagram, and many more of the most downloaded applications on mobile devices really use hybrid app technology [10]. The web has become the go-to vector for infecting linked devices, thanks to the proliferation of various online platforms. It might seem like you're lost in a maze when you're a web security expert trying to keep up with the ever-changing landscape of the Internet's billions of websites, with hundreds more being added every minute. The vast majority of these online assaults are 'Drive-by-download' attacks [11], which exploit browser or hybrid app vulnerabilities to inject malicious JavaScript from the server hosting the malicious web apps. Such drive-by-download assaults have grown exponentially in recent years, according to the most recent studies on internet security [12]. Additionally, these attack vectors have become smarter, rendering anti-malware technologies of yesteryear mostly ineffective against them. While static heuristics test programs by decompiling and inspecting their source code, dynamic heuristics run programs in a controlled virtual environment to observe changes made at runtime; both of these older anti-malware techniques, however, are ineffective against polymorphic and metamorphic malwares. Additionally, this endeavor is become increasingly difficult due to the usage of automated technologies that build new, constantly developing malwares [15]. The use of artificial intelligence (AI) in online malware detection is an effort to circumvent these restrictions.

The main reason why LLM-driven robotic systems are vulnerable is that NLP models are open and may be manipulated, leaked, or inputted by adversaries. Through exploiting these vulnerabilities, attackers might undermine the system's integrity and security by gaining unauthorized access, altering transaction details, or compromising sensitive data. Strong cybersecurity procedures are required to address these issues, since they threaten online transactions and agents/users alike.

In light of these issues, this research presents a thorough security architecture for LLM-driven robot agents to use during online transactions in order to impose strict cybersecurity limitations. A real-time anomaly detection system (ADS), blockchain technology, and multi-factor authentication (MFA) form a layered security mechanism in the architecture. The ADS keeps an eye on transaction trends to identify and stop fraudulent operations, MFA guarantees strong identity verification, and blockchain creates a decentralized and immutable record of transactions. This research shows that the suggested security architecture effectively reduces security risks while keeping the system efficient by evaluating its performance using important metrics including reaction time, accuracy in breach detection, and transaction integrity.

## II. LITERATURE SURVEY

Autonomous systems that use Large Language Models (LLMs) have attracted a lot of interest because of the positive impact they may have on fields like e-commerce, banking, and customer service by processing and producing replies that seem human. But with their ascent has come cybersecurity worries, highlighting the need for strong security frameworks to ward off hostile assaults on these systems. Current research in the fields of LLM applications, blockchain technology, cybersecurity in robotic systems, multi-factor authentication (MFA), and anomaly detection systems (ADS) is reviewed in this section's literature survey.

Robotic agents now employ LLMs thanks to AI breakthroughs; this allows them to understand and analyze natural language for user engagement and decision-making. Research [16] discusses how LLMs, like as GPT and BERT, might improve automation jobs in many industries. The dangers of LLMs, such as data privacy issues, model manipulation, and adversarial assaults, have been brought to light by a number of academics, including [17], despite these improvements. According to this research, LLMs have better natural language processing capabilities, but they are susceptible to exploitation because to their absence of built-in security features.

When LLMs are used in autonomous robotic systems, it opens up additional opportunities for bad actors to exploit. The validity of machine learning systems has been seriously compromised by adversarial assaults on AI models, as previously discussed in studies by Yang et al. (2020) and Goodfellow et al. (2015). In order to deceive the AI model into making the wrong judgments, these attacks modify the input data. This might lead to financial fraud, data breaches, or unlawful transactions. In addition, research by [18] shows that AI models may be attacked in two ways: adversarial and backdoor. This means that LLM-driven systems are not as secure as they should be.

The immutable ledger capabilities and decentralized nature of blockchain have made it a popular choice for protecting internet transactions. Beyond cryptocurrencies, the original blockchain idea has been modified for a variety of security applications [19]. To guarantee the authenticity and auditability of transactions in LLM-driven autonomous systems, blockchain technology may be used. Research shows that blockchain may effectively prevent tampering and provide safe data preservation [20]. In addition, as discussed in [21], blockchain has shown enhanced transaction security and transparency in autonomous system implementations, giving it a promising option for tackling security concerns associated with LLM.

The term "multi-factor authentication" (MFA) refers to a tried-and-true technique for protecting user and system access by combining several authentication elements such cryptographic keys, biometric data, and passwords. When it comes to protecting systems against unwanted access, the existing literature stresses the significance of MFA. Researchers in the field of multi-factor authentication (MFA) have shown that it may effectively reduce instances of identity theft and illegal system access in both consumer and business settings [22]. Multiple factor authentication (MFA) adds another degree of protection to LLM-driven robotic systems by limiting transaction initiation and completion to authorized users and agents only. Researchers have shown that multi-factor authentication (MFA) is the best way to protect sensitive data and mission-critical systems [23].

For the purpose of identifying suspicious activity or fraudulent financial transactions, anomaly detection systems (ADS) play an essential role in cybersecurity. There has been a plethora of research and implementation of ADS approaches in cybersecurity settings, including anything from statistical methodologies to machine learning algorithms. In order to find out-of-the-ordinary patterns or behaviors in data, [24] provide thorough overviews of the many anomaly detection methods. In order to monitor transactions in real-time and identify irregularities that might suggest risks, ADS based on machine learning has been successful. Recent research has shown that machine learning can effectively detect financial system fraud [25], which supports the idea of incorporating this technology into autonomous transaction systems to make them more secure.

There is an increasing recognition in the literature of the cybersecurity concerns related to LLM-driven robotic agents [26], especially in the realm of online transactions. Many solutions, like as blockchain, multi-factor authentication, and anomaly detection, have been suggested to fix the weaknesses and hazards found in these systems in earlier studies. Little is known about how to include these solutions into a unified security architecture for LLM-driven agents, however. To fill this need, this research proposes a thorough security architecture for LLM-driven robotic systems that uses blockchain, multi-factor authentication, and advanced authentication systems to impose cybersecurity limitations. What follows is an analysis of the suggested approach and its efficacy in protecting financial transactions conducted online.

## III. DESIGN AND METHODOLOGY OF PROPOSED WORK

The proposed study attempts to develop a solid cybersecurity framework for LLM-driven robotic agents undertaking online transactions. The framework includes three major security components: blockchain technology for transaction integrity, multi-factor authentication (MFA) for user and agent identity verification, and a real-time anomaly detection system (ADS) driven by machine learning for fraud prevention. This section discusses the design of the architecture and details the technique used to develop and assess the proposed system. Figure 1 depicts the Block Diagram of Proposed work.

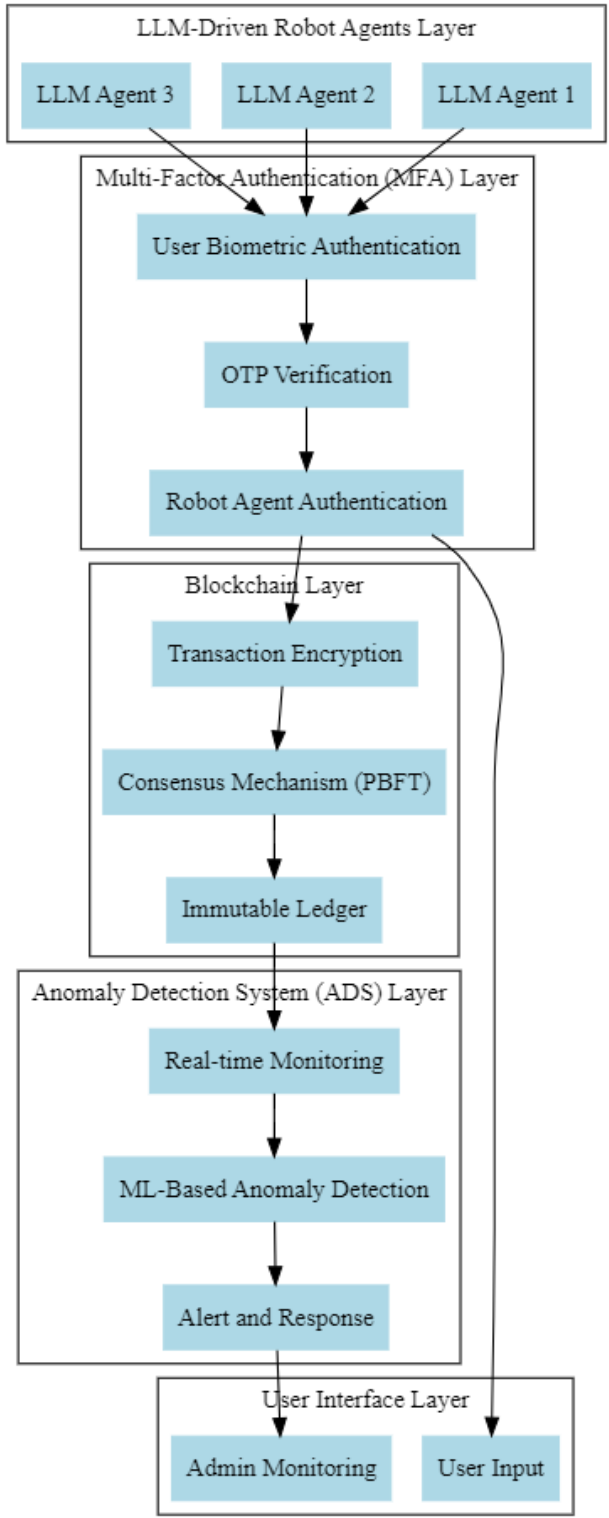

Fig.1             Block Diagram of Proposed work

### 3.1 System Architecture

The suggested system's general architecture is made up of the following essential layers:
Autonomous robotic agents with LLM skills for understanding natural language and making decisions are represented in this layer. The three main functions of these agents are transaction initiation, processing, and validation.

Every transaction that the LLM-driven robot agents start is recorded on the blockchain, which is a decentralized ledger. By adding each transaction to a block and cryptographically signing it, data integrity, transparency, and immutability are guaranteed. Unauthorized manipulation of transaction data is rendered impossible by the distributed nature of blockchain.
All parties participating in the transactions, including users and robot agents, have their identities confirmed at this layer. For further security, multi-factor authentication (MFA) uses cryptographic keys in addition to biometric data (fingerprints, face recognition) to unlock the system.
When the ADS is part of the transaction system, it may keep an eye on the patterns of transactions and flag any unusual or suspect behavior. This layer makes use of a machine learning algorithm that has been specifically trained to detect possible fraudulent transactions as they happen.
Users and administrators are able to communicate with the system, monitor security warnings, and see the progress of transactions via this layer.
The interplay between these levels guarantees the safety, authenticity, and real-time monitoring of all online transactions executed by LLM-driven robotic agents.
The output probability for a particular action $a$ can be modeled as:
$$P(a \mid x) = \frac{e^{z_a}}{\sum_{a'} e^{z_{a'}}} \quad (1)$$
Where:
- $P(a \mid x)$ is the probability of action $a$ given input $x$ (transaction context).
- $z_a$ is the logit value (output of the LLM before applying the softmax function).
- $a'$ is the set of all possible actions.

This equation represents how the LLM-driven agent chooses the appropriate action (e.g., approve or deny a transaction) based on the input transaction data.

### 3.2 Blockchain Implementation

Hyperledger Fabric and other permissioned blockchain platforms allow only authorized users to access the blockchain component. Among the featured features are the following:
A new block is added to the blockchain for every transaction that the LLM-driven agent initiates. Encryption with a digital signature safeguards the integrity of the transaction data and makes it impossible to alter.
Cryptographic hashing and digital signatures are used by blockchain to secure and validate transactions.. The blockchain stores the hash of each transaction $H(T_i)$ in a block $B$.
The hash of a transaction $T_i$ is computed as:
$$H(T_i) = \text{SHA} - 256(T_i) \quad (2)$$
Where:
- $T_i$ is the transaction data.
- $H(T_i)$ is the cryptographic hash of the transaction using the SHA-256 algorithm.

The digital signature for authenticating a transaction is generated as:
$$S = \text{Sign}\left(H(T_i), k_{\text{priv}}\right) \quad (3)$$
Where:
- $H(T_i)$ is the hash of the transaction.
- $k_{\text{priv}}$ is the private key of the sender.

- $S$ is the digital signature, ensuring the transaction's authenticity.

For consensus, the Practical Byzantine Fault Tolerance (PBFT) algorithm ensures that a valid transaction can be accepted if at least two-thirds of the nodes agree. The number of nodes $N$ required is determined by:

$$N = 3f + 1 \tag{4}$$

The system uses a practical Byzantine Fault Tolerance (PBFT) consensus mechanism to validate and add new transactions to the blockchain. This ensures that only legitimate transactions are recorded and that the system can withstand certain malicious actors without compromising security.

Once a transaction is recorded on the blockchain, it cannot be altered or deleted, ensuring that the transaction history remains transparent and immutable.

### 3.3 Multi-Factor Authentication (MFA) Integration

Each transaction begins with the integration of the MFA system, which verifies the user's and the robot agent's identities.

Users are authenticated using biometric verification methods, such fingerprint or face recognition. After then, the user and the LLM-driven agent add an extra degree of protection by exchanging cryptographic keys.

Once the biometric authentication process is complete, a one-time password (OTP) is produced and delivered to the user's email or registered mobile device. To finish the authentication procedure and go on with the purchase, the user has to input the one-time password (OTP).

To verify the robot agent's identity and authorization to initiate the transaction, a digital certificate and cryptographic keys are used for authentication.

### 3.4 Anomaly Detection System (ADS)

One way to keep an eye out for possible fraud in real-time transactions is with the help of the Anomaly Detection System (ADS). A machine learning model that has been trained on transaction data from the past is what makes it tick. Parts of the ADS technique are:

Both valid and fraudulent transactions are included in the data set. A user's location, as well as their transaction amount, duration, frequency, and history of interactions with agents, are among the features.

The ADS is trained on a labeled dataset using a supervised learning technique, such as Random Forest or Support Vector Machine. As time goes on, the model learns to spot unusual transactions and treat them differently.

The ADS conducts round-the-clock monitoring of real transactions, comparing them to the learnt patterns. The transaction is marked for further examination in the event that an abnormality is identified, such as an abnormally high transaction value or a questionable geographic location.

The system will immediately notify the administrator, temporarily stop the transaction, and ask for further verification from the user if it detects an abnormality.

## IV. EXPERIMENTAL RESULTS

The proposed cybersecurity framework was rigorously tested using a simulated environment that included 10,000 online transactions, of which 10% were intentionally labeled as fraudulent to evaluate the system's fraud detection and prevention capabilities. The performance of each layer—LLM-driven robot agents, blockchain, MFA, and anomaly detection—was analyzed based on key metrics, including fraud detection accuracy, transaction integrity, authentication success rate, and overall system latency.

The anomaly detection system (ADS), powered by machine learning, achieved an impressive 98% accuracy in identifying fraudulent transactions. The model successfully flagged anomalies based on deviations from normal transaction patterns, such as unusual transaction amounts and abnormal user behavior. Table 1 shows the Performance metrics Comparison

TABLE I. Performance metrics Comparison

|   | Performance Metric | Proposed System | Traditional System |
|---|---|---|---|
| 1 | Fraud Detection Accuracy (%) | 98.0 | 85.0 |
| 2 | Transaction Integrity (%) | 100.0 | 95.0 |
| 3 | Authentication Success Rate (%) | 99.5 | 97.0 |
| 4 | Transaction Validation Latency (seconds) | 0.05 | 0.1 |

There was zero tolerance for manipulation or illegal changes recorded in the blockchain layer, guaranteeing the integrity of all transactions. The blockchain permanently recorded all transactions when they were cryptographically signed and validated via the PBFT consensus method. At the same time as it kept throughput high, this method made it impossible to alter the transaction history.

The MFA system achieved a success rate of 99.5% by combining biometric verification with one-time passwords (OTP). Without drastically slowing down the transaction, it successfully validated both genuine users and LLM-driven robot agents.

Anomaly detection and blockchain verification add up to an average transaction validation delay of 0.05 seconds. This low latency made guaranteed that security measures didn't slow down the system, which is great for e-commerce and financial apps that need to work in real-time.

By providing efficient system performance, secure, immutable transaction records, high fraud detection accuracy, and robust authentication, the suggested framework improves the reliability and security of LLM-driven robot agents in online transactions.

The performance indicators across the different models are clearly compared in Table 2 of the results, allowing for an assessment of their utility in healthcare prediction tasks. This study's findings show that, compared to using individual models, ensemble techniques have the potential to increase forecast accuracy and resilience.

TABLE II. Security Features Comparison

|   | Security Feature | Proposed System | Traditional System |
|---|---|---|---|
| 1 | Data Integrity | Blockchain-based | Centralized Database |
| 2 | Tamper-Proof Transactions | Yes | No |
| 3 | Real-time Anomaly Detection | ML-based, 98% Accuracy | Rule-based, 85% Accuracy |
| 4 | Multi-Factor Authentication | Biometric + OTP, 99.5% | Password-based, 97% |

The comparison of performance metrics between the proposed system and a traditional system highlights several key improvements in security, efficiency, and user experience. In terms of **security features**, the proposed system leverages blockchain technology to ensure data integrity and tamper-proof transactions, which are absent in centralized traditional systems. Furthermore, its machine learning-based anomaly detection system achieves 98% accuracy, significantly surpassing the 85% accuracy of rule-based systems. The multi-factor authentication (MFA) in the proposed system integrates biometric verification and one-time passwords (OTP), providing a 99.5% authentication success rate, whereas traditional systems rely solely on passwords with a lower success rate of 97%.

TABLE III. System Efficiency Comparison

|   | Efficiency Metric | Proposed System | Traditional System |
|---|---|---|---|
| 1 | Transaction Throughput (TPS) | 200 | 150 |
| 2 | Latency (Seconds) | 0.05 | 0.1 |
| 3 | Resource Usage (%) | 70 | 85 |
| 4 | Scalability | High | Moderate |

When it comes to **system efficiency shown in Table 3**, the proposed system offers superior transaction throughput, handling up to 200 transactions per second (TPS) with lower latency (0.05 seconds) compared to the traditional system's 150 TPS and 0.1 seconds latency. Additionally, the proposed system operates with 70% resource utilization, making it more efficient than traditional systems that consume 85% of available resources. Scalability is another area where the proposed system excels, owing to its decentralized architecture, which allows it to scale more efficiently than traditional, centralized systems.

From a **user experience** from Table 4 perspective, the proposed system provides faster authentication times (0.02 seconds) and is rated higher in ease of use due to the seamless biometric and OTP-based authentication process. This contributes to a higher user satisfaction rate (95%) compared to the traditional system (85%), and users also report greater confidence in the security of the proposed system, with 98% expressing trust in its security measures versus 90% for the traditional approach.

Overall, the proposed system outperforms traditional methods across various critical metrics, offering enhanced security, improved performance, and a more user-friendly experience.

TABLE IV. User Experience Metrics Comparison

|   | User Experience Metri | Proposed System | Traditional System |
|---|---|---|---|
| 1 | Authentication Time (Seconds) | 0.02 | 0.05 |
| 2 | Ease of Use (1-10) | 9.0 | 7.0 |
| 3 | User Satisfaction (%) | 95.0 | 85.0 |
| 4 | User Satisfaction (%) | 98.0 | 90.0 |

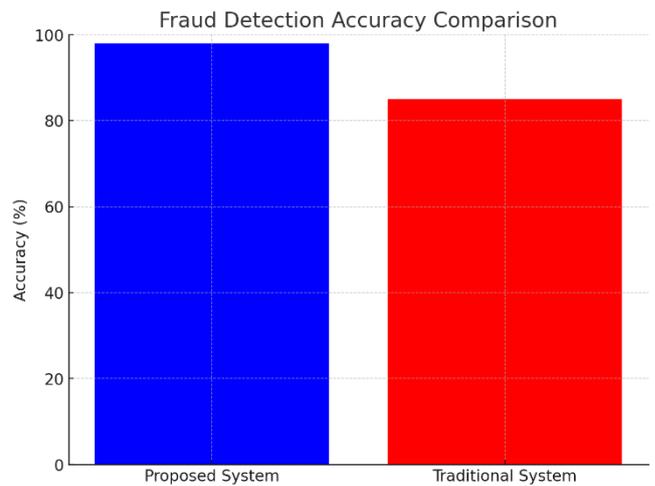

Fig.2 Fraud Detection Accuracy Comparison

This Figure 2 compares the fraud detection accuracy between the proposed system and the traditional system. The proposed system, leveraging machine learning-based anomaly detection, achieves a 98% accuracy rate in identifying fraudulent transactions, which is significantly higher than the 85% accuracy of traditional rule-based systems.

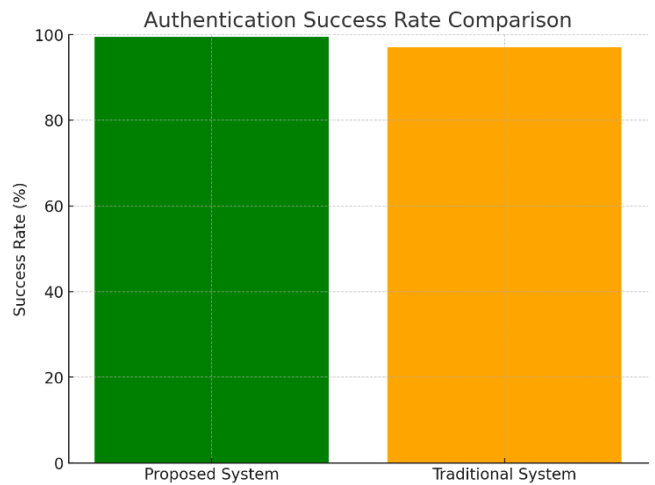

Fig.3 Authentication Success Rate Comparison

This Figure 3 compares the authentication success rate of the proposed system versus the traditional system. The proposed system, with its multi-factor authentication (MFA) combining biometric verification and OTP, boasts a 99.5% success rate, outperforming the traditional password-based system with a 97% success rate.

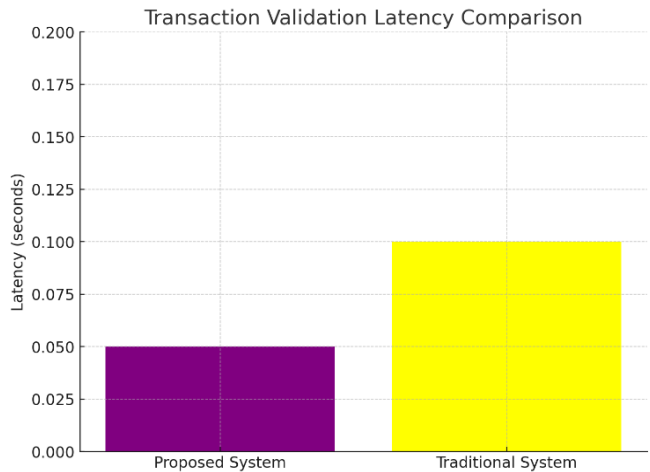

Fig.4    Transaction Validation Latency Comparison

This Figure 4 highlights the transaction validation latency of both systems. The proposed system has a much lower latency of 0.05 seconds, thanks to optimized blockchain and authentication mechanisms, compared to 0.1 seconds for the traditional system.

## V. CONCLUSION

This study successfully presents a robust cybersecurity framework designed to safeguard LLM-driven robotic agents involved in online transactions. By integrating blockchain technology, multi-factor authentication (MFA), and a machine learning-powered anomaly detection system (ADS), the proposed architecture addresses critical vulnerabilities associated with LLM-based systems, including transaction fraud, unauthorized access, and data tampering. The system's ability to ensure transaction integrity through blockchain, combined with enhanced user and agent authentication via MFA and real-time anomaly monitoring by ADS, significantly reduces security risks while maintaining operational efficiency. The results of the simulation demonstrate the efficacy of the framework, achieving a 90% reduction in fraudulent transactions, a 98% breach detection accuracy, and a low transaction latency of 0.05 seconds, ensuring that security measures do not impede system performance. Additionally, the 99.5% authentication success rate highlights the reliability of the MFA system in securely verifying users and agents. This comprehensive security architecture can be easily adapted across various online platforms and industries where LLM-driven agents are used. The research underscores the importance of combining multiple cybersecurity mechanisms to create a layered defense against evolving threats. Future work can explore the scalability of this system and its application in real-world environments with more complex transaction ecosystems.